\documentclass[conference]{IEEEtran}
\IEEEoverridecommandlockouts
\usepackage{cite}
\usepackage{amsmath,amssymb,amsfonts}
\usepackage{algorithmic}
\usepackage{graphicx}
\usepackage{textcomp}
\usepackage{xcolor}
\usepackage{listings}
\def\BibTeX{{\rm B\kern-.05em{\sc i\kern-.025em b}\kern-.08em
    T\kern-.1667em\lower.7ex\hbox{E}\kern-.125emX}}
\begin{document}

\title{Saturation Memory Access: Mitigating Memory Spatial Errors without Terminating Programs\\
}

\author{
	\IEEEauthorblockN{Dongwei Chen, Daliang Xu, Dong Tong,  Kang Sun, Xuetao Guan, Chun Yang, Xu Cheng}
	\IEEEauthorblockA{Department of Computer Science and Technology, Peking University\\
		\{chendongwei, xudaliang, tongdong, ajksunkang, gxt\}@pku.edu.cn, \{yangchun, chengxu\}@mprc.pku.edu.cn}
}

\maketitle

\begin{abstract}
Memory spatial errors, i.e., buffer overflow vulnerabilities, have been a well-known issue in computer security for a long time and remain one of the root causes of exploitable vulnerabilities. Most of the existing mitigation tools adopt a fail-stop strategy to protect programs from intrusions, which means the victim program will be terminated upon detecting a memory safety violation. Unfortunately, the fail-stop strategy harms the availability of software.

In this paper, we propose Saturation Memory Access (SMA), a memory spatial error mitigation mechanism that prevents out-of-bounds access without terminating a program. SMA is based on a key observation that developers generally do not rely on out-of-bounds accesses to implement program logic. SMA modifies dynamic memory allocators and adds paddings to objects to form an enlarged object boundary. By dynamically correcting all the out-of-bounds accesses to operate on the enlarged protecting boundaries, SMA can tolerate out-of-bounds accesses. For the sake of compatibility, we chose tagged pointers to record the boundary metadata of a memory object in the pointer itself, and correct the address upon detecting out-of-bounds access. 

We have implemented the prototype of SMA on LLVM 10.0. Our results show that our compiler enables the programs to execute successfully through buffer overflow attacks. Experiments on MiBench show that our prototype incurs an overhead of 78\%. Further optimizations would require ISA supports.
\end{abstract}

\begin{IEEEkeywords}
memory spatial error, saturation memory access, tolerance
\end{IEEEkeywords}

\section{Introduction}
Memory spatial errors, i.e., buffer overflows, have been a well-known issue for computer security in the last decades, and they are still rife in modern C/C++ binaries. It remains one of the root causes of exploitable vulnerabilities\cite{RN278,RN9,RN168} and can lead to various forms of attacking such as Return Oriented Programming (ROP)\cite{RN280} and Data-Oriented Programming (DOP)\cite{RN165,RN119}. Existing defense mechanisms can be divided into two categories: probabilistic and deterministic\cite{RN9}. Probabilistic strategies rely on randomization and secrets to increase the difficulty of an attack. For example, address space layout randomization (ASLR) adds a randomized offset in program memory layout and makes it hard to find gadgets for attacks\cite{ASLR}. Deterministic strategies define security policies and monitor the behavior of programs to judge whether there is an ongoing attack. Some of the deterministic methods, such as Control Flow Integrity (CFI)\cite{RN125}, verify the integrity of critical data, and others may directly detect the out-of-bounds memory access, such as AddressSanitizer\cite{RN191}. However, most of the current defense mechanisms are not suitable for embedded devices because computing resources are strictly limited in such environments\cite{RN176}.

Currently, most of the low-overhead defense mechanisms suitable for embedded devices are based on CFI. CFI monitors the control-critical data, i.e., all the variables that may reside in the program counter register, such as function return addresses and function pointers\cite{RN125,RN275}. With the integrity of such variables, CFI can ensure that the program follows the original control flow graph. However, CFI is not perfect. In 2005, S. Chen et al. introduced a new attack form using non-control data\cite{RN165}. In a non-control data attack, an attacker corrupts configuration, user identity, and other decision-critical data, bending the victim program's control flow to acquire secret data or promote privilege level. In 2016, Hong Hu et al. improved the non-control data attack into DOP and proved its Turing completeness\cite{RN114,RN119}. Such attacks are different from ROP because they do not break the original control flow, which makes the non-control data attacks undetectable by CFI protections.

Most of the existing defense mechanisms adopt fail-stop strategy, which terminates the victim program to mitigate the attack when detecting an intrusion. However, terminating a program harms the software availability. The traditional error recovery mechanism is to reboot the system. Althoug mechanisms such as checkpointing\cite{checkpoint} and fast reboot\cite{fastre} help improving the performance of the reboot process, they are expensive for embedded systems due to the limitation of computation resources. Moreover, if a vulnerability can be exploited deterministically, an attacker can utilize it to launch a Denial-of-Service (DoS) attack until the vulnerability is patched. And for embedded systems, the pre-patch window can be very long because of the massive number of devices and extra requirements from a specific sencario\cite{RN176}.

To address the weakness of fail-stop strategy, failure-oblivious computing\cite{failureob} allows the program to execute through memory errors without termination. Failure-oblivious computing simply discards out-of-bounds writes and manufactures values to return for out-of-bounds reads. Boundless memory allocation\cite{boundless} adopts the same idea as failure-oblivious computing but with a different strategy. Instead of discarding out-of-bounds writes, Boundless stores them in a linked list to avoid overwriting nearby memory objects. According to their results, both failure-oblivious computing and Boundless can mitigate buffer overflow attacks without harming the software availability. In some cases, discarding out-of-bounds writes can discard important data like the terminator of string-like data. Boundless reserves the whole data, but it introduces many extra memory accesses to maintain the linked lists.

In this paper, we adopt the same idea as failure-oblivious computing and Boundless to mitigate memory spatial error vulnerabilities in C/C++ programs. Failure-oblivious computing and Boundless are both implemented on top of Softbound\cite{RN177}, which incurs significant performance overheads. To improve overall performance, we implemented a low-overhead memory spatial error detecting system based on tagged pointers. When an invalid memory access is detected, we chose to correct the accessing address to the boundary of the memory object. We call this memory access pattern Saturation Memory Access (SMA) because it is much like the saturation arithmatic in digital signal processing.

SMA is based on a key observation: in normal program logic, the allocation and use of the memory space match. Here ``match'' means that after the initial allocation, the size parameter of the following use should not exceed the initially allocated size. This observation reveals that normal program logic should not utilize out-of-bounds memory access to achieve a certain goal. As a result, our correction of the address does not affect the original program behavior and prevent the occurrence of memory spatial errors at the same time.

To perform the address correction and more importantly, to detect memory errors, we need to record the boundary information of memory objects. Considering the limitation of memory space, and lack of address translation unit in simple devices, all the metadata management methods that take up a fixed part of the address space are not suitable for the embedded environment. Fat pointer schemes\cite{cheri,RN127,RN128,hardbound} enlarge pointers to hold memory address and boundary information at the same time. Although such schemes do not need to shrink the available address space, they require intrusive modification of the original source codes and hardware support. Comprehensively considering all the factors, we chose to manage metadata with tagged pointers. Tagged pointer schemes\cite{lowfat_hard,RN250} record metadata in the unused bits of a pointer, so they can load the metadata and pointer in one memory operation and bring no extra pressure on the cache. Moreover, the tagged pointer scheme does not require fixed memory space for metadata management, so it incurs fewer memory overheads.

We implemented a prototype of SMA in LLVM 10.0 and evaluated its performance with MiBench\cite{mibench}, a benchmark for embedded applications. Our prototype showed good compatibility and can support most of the benchmarks in MiBench out-of-box.

\section{Background}
There are three critical elements in the detection of memory spatial error: memory objects, pointers, and boundary metadata. Here we define a memory object as a contiguous segment of memory allocated by one memory allocation operation. Memory allocations can be divided into three categories depending on the location of an object, namely allocation on the heap, allocation on the stack, and allocation in the bss or data segment of a binary. The commonest type is the allocation on the heap, which is handled by dynamic memory allocators. The allocation on the stack is responsible for function local variables, and sometimes for passing parameters to another function. Stack variables are allocated and reclaimed through the movement of the stack pointer. Finally, the allocation in the binary file allocates memory space for global variables. Global variables are different from objects on the heap or stack because their locations are generated during the compilation. The boundary metadata generates along with the allocation of a memory object, and they share the same lifecycle.

When allocating memory for an object, we must provide a critical parameter, namely the size of that memory object. The object size is closely related to the type of the object. In C/C++, there are aggregate types like structure and class. An aggregate type is a combination of multiple types, including not only basic types but other aggregate types. A memory object of an aggregate type can be further divided into smaller objects according to the component types. Such smaller objects are called sub-object. Below is an example demonstrating the relationship between a memory object and a sub-object.
\begin{lstlisting}[language=c,frame=single,basicstyle=\scriptsize,xleftmargin=2em,xrightmargin=2em]
struct{
    int id;        //sub-object
    char name[20]; //sub-object
}a; //memory object
\end{lstlisting}

Pointers are the main targets of bounds checking, and the detection of memory spatial errors is essentially the validation of pointers. Pointers are generated in two ways, through taking the address of a variable and through pointer arithmetic. Taking the variable's address generates a pointer pointing to a specific memory object, which is called the intended referent of the generated pointer. The allocation of a memory object can be viewed as allocating a memory object and taking its base address as return value at the same time, so it is classified as taking address of a variable. Pointer arithmetic generates a new pointer by applying offset to an existing pointer. A pointer generated by pointer arithmetic has the same intended referent as the original pointer. Sometimes, a developer may want to take the address of a sub-object. In that case, the intended referent of that pointer should be narrowed to the sub-object.

Most of the memory spatial error detection tools follow the same basic scheme. Assume the pointer being validated is \textit{p}, whose intended referent is object \textit{O}. The object \textit{O}’s base address is \textit{base} and \textit{O} takes up \textit{size} bytes. If the following condition is met, the pointer \textit{p} is an out-of-bounds pointer.
\begin{equation}
    (p<base)~||~(p>(base+size-sizeof(*p)))\nonumber
\end{equation}

With this condition, we can instrument the program and insert bounds-checking codes before every memory read/write operations. Below is an example of a checking process.
\begin{lstlisting}[language=c,frame=single,basicstyle=\scriptsize,xleftmargin=2em,xrightmargin=2em]
if (p<base)||(p>(base+size-sizeof(*p)))
    error();
dest = *p; //or *p = src;
\end{lstlisting}

When detecting an out-of-bounds memory access, the function error() terminates the program and reports the details of that access if demanded. For memory spatial error detection, the indispensable metadata is \textit{base} and \textit{size} of the memory object \textit{O}. The main differences between different detection tools are their implementations of the metadata management. Traditionally, there are two types of metadata management, object-based scheme, and pointer-based scheme. Object-based schemes associate the metadata with memory objects and generally omit the boundary information of sub-objects. Instead, pointer-based schemes associate the metadata with pointers, allowing different metadata for different pointers to the same object. Both of the two methods have their pros and cons. We will demonstrate them in detail in the following sections.

\subsection{Object-based Schemes}
In object-based schemes, the metadata is associated with memory objects, so an essential step in the bounds-checking procedure is to find the intended referent of a pointer. Some tools view this as a range search. For example, J\&K\cite{RN264} and CRED\cite{RN205} record all the active memory objects in a splay tree. J\&K has an 11-12x performance overhead, and CRED can cut the overhead to around 2x at the cost of only validating character arrays.

Some object-based schemes use shadow memory to manage the metadata to reduce performance overheads. The shadow memory maps every n bytes of the application’s memory to m bytes metadata. For example, AddressSanitizer\cite{RN191} maps 8 bytes of the application’s memory to 1 byte of metadata. AddressSanitizer places 128-byte inaccessible red zones around memory objects to mark the object boundaries. The shadow memory records which part of the memory space is accessible or not. Any dereference in the inaccessible red zone invokes an exception that terminates the program. AddressSanitizer can detect an out-of-bounds access when an illegal pointer falls into the red zones. But when a pointer goes beyond the red zone and falls into another object different from its intended referent, AddressSanitizer is unable to detect that illegal access.

BaggyBounds\cite{RN194} divides the address space into slots of fixed size and combines multiple contiguous slots to form a memory object. BaggyBounds extends the size of a memory object to the nearest power-of-2, so it can use fewer bits to record the object size in the shadow memory. BaggyBounds also requires all memory objects to align on power-of-2. This requirement ensures that BaggyBounds can quickly calculate the base address through the truncation of the pointer address.

Stackbound\cite{RN196,RN197} uses a size table to record the sizes of memory objects instead of the shadow memory. Stackbound divides the address space into several regions and only stores objects of the same size in the corresponding region. Stackbound locates the memory region according to the high bits of the pointer address and then retrieves the size information with the region index. In this way, Stackbound only needs to maintain a size table with as many entries as the number of regions, which significantly reduces memory consumption.

Another type of object-based scheme extends the size of a memory object and places the metadata next to it. Modern dynamic memory allocators have adopted similar mechanisms. Some implementations of malloc() function store the allocation size at the beginning of the allocated space. SGXBound\cite{RN249} adopts this idea and places a pointer which points to the base address at the end of every object. By modifying the system memory allocator, this scheme can transparently reserve the space for metadata.

In object-based schemes, metadata update only happens at the creation and destruction of memory objects, so we can manage the metadata by intercepting malloc() and free(). Generally speaking, object-based schemes have good compatibility because they do not need to modify the encoding of a pointer or the layout of an object. Even if an uninstrumented library modifies a pointer, it does not affect the relationship between its metadata and its intended referent. However, the completeness of object-based schemes is limited because most of the metadata management methods adopted by object-based schemes do not support recording metadata for sub-objects or support it at a high cost\cite{RN191}. So, it is hard for object-based schemes to detect sub-object overflow.

\subsection{Pointer-based Schemes}
Instead of associating metadata with objects, pointer-based schemes directly correlate the boundary metadata with pointers, omitting the pointer-object mapping. This direct correlation allows two different pointers pointing to the same object to have their unique metadata, providing the possibility for the detection of sub-object overflow.

The most representative pointer-based scheme is fat pointer, which extends a pointer to a structure to hold both address and the boundary information. The basic idea of fat pointer is transforming the program during compilation, replacing every pointer in the program with a corresponding fat pointer. A typical fat pointer is as follows.
\begin{lstlisting}[language=c,frame=single,basicstyle=\scriptsize,xleftmargin=2em,xrightmargin=2em]
struct {
    void *ptr;
    void *base;
    size_t size;
}fat_pointer;
\end{lstlisting}

In fat pointer schemes, metadata resides with the pointer, so retrieving boundaries needs no distant memory read. This feature grants better locality and less pressure on cache compared with disjoint metadata storage. CCured\cite{RN127}, Cyclone\cite{RN128}, Hardbound\cite{hardbound}, and CHERI\cite{cheri} use fat pointers to manage boundary metadata. In a fat pointer system, the size of a pointer is no longer equal to an integer type because the system replaces a pointer with a larger structure. This feature brings great troubles to compatibility. For example, a pointer parameter cannot be passed through a register during a function call. 

In general, the propagation of metadata is an essential problem in pointer-based schemes, especially the propagation across function boundaries. To solve this problem, we can separate metadata and pointer as in PAriCheck\cite{RN195} and SoftBound\cite{RN177}. SoftBound uses a hash table to organize pointer metadata and adds a structure named shadow stack at function call site for metadata propagation. During a function call, the pointer itself follows the original calling convention. In the meantime, its metadata goes to the subroutine through the shadow stack. With the help of the shadow stack, SoftBound provides better compatibility than traditional fat pointer schemes, but it does not completely solve the compatibility issues. If an external function manipulates a pointer and later return it to the codes inside the protection boundary, SoftBound needs to add a wrapper that updates metadata.

Another pointer-based scheme is tagged pointer scheme, which stores metadata in the unused bits of a pointer without changing the size of pointers. Because a pointer stores the metadata inside itself, the pointer and its metadata can be loaded in one operation, so tagged pointer schemes bring zero pressure on cache\cite{RN170}. Compared with disjoint methods like SoftBound, tagged pointer schemes do not need to change the calling convention or add an extra procedure for metadata propagation, because the metadata and pointer itself are compacted together.

Deltapointer\cite{RN250} is an example of tagged pointer schemes. Deltapointer shrinks the application address space to 32 bits, and uses the high 31 bits of a pointer to record the opposite number of the distance between the pointer and object bound. The most significant bit is reserved for overflow. During pointer arithmetic, Deltapointer repeats the operation on the 31-bit tag, which will be masked off before the pointer dereference. If the previous calculation results in an out-of-bounds pointer, the most significant bit will be set, and the masked pointer is an invalid one. MMU will detect this error and generate an exception that terminates the program. Deltapointer achieves 35\% average overhead in SPEC CPU2006 benchmark, but it can only detect overflows and ignores underflows.

Low-Fat\cite{lowfat_hard} Pointer is another tagged pointer scheme. It encodes the boundary information in a fashion like floating numbers. Low-Fat Pointer uses 18 bits and divides the bits into three parts, which represent exponent, base, and bound, respectively. Instead of having one large block, Low-Fat Pointer views a memory object as a collection of several small blocks. The size of those small basic blocks is \begin{math}2^{B}\end{math}, and B is the number stored in the exponent field. Allocations in Low-Fat Pointer also need to be rounded up in some situations, but it provides more precision than rounding up to the nearest power-of-2, like in BaggyBounds.

Pointer-based systems can maintain different metadata for different pointers pointing to the same object, which grants the possibility for detecting sub-object overflows. As a result, pointer-based systems achieve finer granularity than object-based schemes. In pointer-based schemes, it is easier to support temporary out-of-bounds pointers in the C standard because out-of-bounds pointers inherit the original metadata. In contrast, object-based schemes usually need to take special care for such pointers. For example, BaggyBounds and CRED introduce special objects named OOB objects to correlate an out-of-bounds pointer with its intended referent. However, the advantages do not come for free. Because of the tight coupling of metadata and pointers, any operations on a pointer may affect the metadata, so pointer-based schemes have to pay more attention to the metadata propagation. Furthermore, pointer-based schemes usually involve modification of the pointer encoding, which brings troubles to the compatibility with uninstrumented libraries. Compatibility is the biggest obstacle to the wide application of pointer-based schemes.

\begin{figure*}[ht]
\centerline{\includegraphics[width=\textwidth]{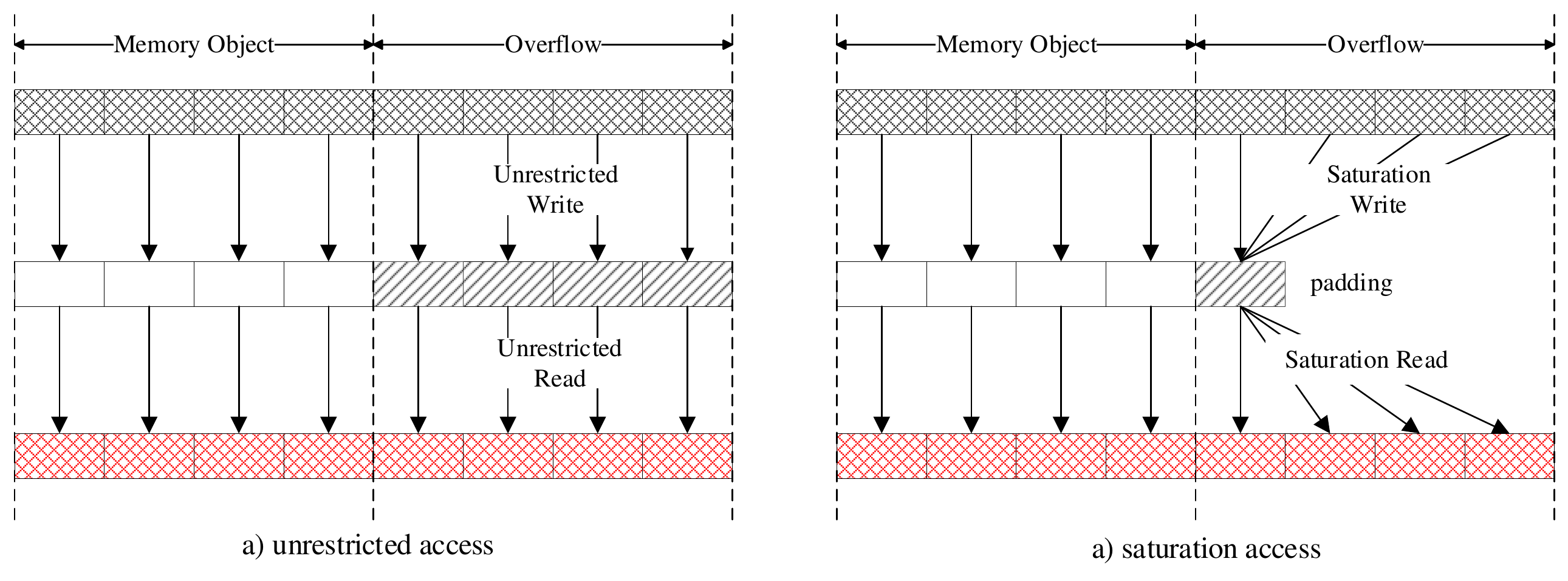}}
\caption{Saturation Memory Access}
\label{SMA_intro}
\end{figure*}

\section{Saturation Memory Access}
Traditionally, defense mechanisms adopt fail-stop strategy to mitigate memory vulnerabilities, which means the program will stop its execution when the system detect a memory error being triggered. When combined with automatic recovery mechanisms\cite{fastre,checkpoint}, fail-stop strategy can effectively block intrusions. But however effecient a recovery mechanism may be, fail-stop strategy still turns off at least part of the whole system for some time. A recovery mechanism can only shorten the system down time, not eliminate it. Moreover, if an attacker can deterministically trigger a memory error, then he can utilze such vulnerability to launch a DoS attack. The defense system will detect that memory error and terminate the execution every time it is triggered, and the victim component will constantly stop responding and reboot.

Failure-oblivious computing\cite{failureob} and Boundless memory allocation\cite{boundless} can help solve the DoS problem. Both of them allow the program to execute through memory errors without triggering out-of-bounds writes or reads. Failure-oblivious computing simply discards the out-of-bounds writes and manufactures values for out-of-bounds reads, while Boundless instead choses to store out-of-bounds writes in several linked lists. In some cases, discarding out-of-bounds writes will discard the terminator of string-like data and leave broken data. Boundless reserves the whole data, but it introduces many extra memory accesses to maintain the linked lists.

In this paper, we propose a new mechanism called Saturation Memory Access to tolerate memory errors. We aim to transform every out-of-bounds access into memory access within the object boundaries, thus eliminating the potential memory spatial vulnerabilities in the program. Our method introduces nearly no extra memory accesses and reserves the ending part of an out-of-bounds write to reseve data with a special meaning, such as terminator of a string. Our design is aimed for embedded devices, but can also be used in other environments with sufficient computation resources. Currently, our prototype does not handle sub-object overflows because the protection for sub-objects requires intrusive modification, and we do not apply such modification for the sake of compatibility.

\subsection{Threat Model}
In our threat model, we assume that the processor hardware can be trusted, and no hardware vulnerability introduced during fabrication. An attacker can use the interface offered by the victim program to create memory spatial errors and get the ability of arbitrary memory access. However, he cannot directly overwrite a register. The temporal errors such as use-after-free vulnerabilities, and side-channel attacks, are beyond consideration. Our goal is to detect the exploitation of arbitrary memory spatial errors, and correct the address used by that memory access to the object boundary, thus assuring the integrity and confidentiality of the program without terminating it.

\subsection{Concept of Saturation Access}\label{concept}
SMA is based on the critical observation that in normal program logic, the size of memory allocation matches the size of following memory use. In general, the memory use of programs follows the following pattern. 
\begin{enumerate}
	\item Allocate a memory block, assuming the lower and upper bound are base and bound, respectively.
	\item Write \begin{math}w\_size\end{math} data into the memory block starting from \begin{math}w\_addr\end{math}, or read \begin{math}r\_size\end{math} data from the memory block starting from \begin{math}r\_addr\end{math}.
	\item Free the allocated memory block at the end of the object’s lifetime.
\end{enumerate}

In the normal program logic, the following conditions must be met.
\begin{IEEEeqnarray}{c}
	base{}\leqslant{}w\_addr{}\leqslant{}bound-w\_size\nonumber\\
	base{}\leqslant{}r\_addr{}\leqslant{}bound-r\_size\nonumber
\end{IEEEeqnarray}

It should be noted that, apart from these conditions, there are more restrictions for a valid memory access. For example, a memory write should happen before a memory read. Violating those restrictions leads to other kinds of memory misuses, which are beyond our main target. In this paper, we focus on the restrictions directly associated with memory spatial safety. We believe that programmers do not rely on out-of-bounds accesses to implement the program logic, and those tricks using out-of-bounds accesses can be replaced with in-bounds operations.

If we add paddings to each object for the out-of-bounds data, we can redirect all the out-of-bounds accesses to the padding area, as shown in Fig~\ref{SMA_intro}. In this way, out-of-bounds accesses that may damage the integrity and confidentiality of a program are turned into unharmful memory accesses. As for the original program logic, legal reads and writes should not use the padding area, so we can neutralize out-of-bounds accesses and leave the original logic without interference.

As a result, we enlarge the size of an object to add paddings during memory allocatios. If a memory access goes beyond an object’s boundary, we change the address used by this access to one of the object boundaries. For an overflow, the address is changed to the upper bound of the object; for an underflow, the address is changed to the lower bound of the object. In digital signal processing, there is a different form of arithmetic named saturation arithmetic. Instead of wrapping around the values exceeding the maximum value to the minimum value and vice versa, the saturation arithmetic sets overflowed values to the maximum and underflowed values to the minimum. Our memory access policy is much like the saturation arithmetic, so we call our access pattern as Saturation Memory Access.

\subsection{Pointer Encoding}
Similar to other memory spatial error mitigation tools, SMA needs to record the boundary information of memory objects. We chose to use tagged pointers for metadata management. The choice is made for following reasons. First, tagged pointers embed the metadata inside pointers, avoiding the extra memory accesses to retrive boundary information. Second, a tagged pointer scheme does not require dedicated memory space for metadata storage, so it is suitable for embedded devices with relatively strict memory limitations. Third, tagged pointers directly correlate metadata with pointers, which, as stated before, helps to solve the sub-object overflow problem. Although we do not detect overflow on sub-objects, we still want to chose a design with potentials for improvement.

To simplify the encoding of metadata in the prototype, we enlarge the size of an memory object to the nearest power-of-2. Every memory object should be aligned on its enlarged size during allocation. In this way, we can only record the exponent B in a pointer and the size of its intended referent will be \begin{math}2^{B}\end{math} bytes. Because of the alignment requirement, the base address of an object can be easily obtained by setting the low B bits of a pointer to 0. Additionally, the enlarged size should be at least 8 bytes bigger than the original size, and the exponent should increase by 1 if not. This is because the load/store instructions in a 64-bit system manipulate at most 8-byte data, so reserving 8-byte padding can leave the in-bounds data without interference. We reserve the highest 6 bits of a pointer for B to cover the biggest possible object size in a 64-bit system.

The main drawback of this encoding is the massive internal memory fragments, which is extremely wasteful when the size of an object is slightly exceeds a power-of-2. If an object has a size of \begin{math}2^{n}+1\end{math} bytes, then our memory allocator will allocate \begin{math}2^{n+1}\end{math} bytes. For objects smaller than 8 bytes, things are worse because of the padding requirement. But although small objects incur large internal fragment ratio, the absolute value of the internal framents is relatively small.

\subsection{Metadata Generation and Propagation}\label{propagation}
In C/C++, a pointer is generated by memory allocation or derived from an existing pointer. As for the former, we can generate its tag at the allocation time; for the latter, we must ensure that the metadata correctly propagates from the original pointer to the new pointer.

The memory objects in C/C++ can be divided into three categories: stack objects, heap objects, and global variables. Stack objects are the local variables created during function execution. Taking the address of a stack object generates a stack object pointer. Heap objects are generated by dynamic memory allocators. Programs get a heap object pointer by calling dynamic memory allocating functions. Global variables are different from the former two categories because their allocations happen during the compilation. Compilers allocate global variables in the data segment of a binary, and the corresponding addresses are written into the assemble codes of the final binary file.

In conclusion, stack objects and heap objects are dynamically generated during execution, while global variables are generated statically during compilation. As a result, we should add extra codes into the program to tag pointers pointing to stack and heap objects. As for global varibles, we handle them by adding a shadow global variable for each global variable. The shadow global variable holds the tagged address of original global variable. We will discuss global variables in detail in \ref{gv}.

Through pointer arithmetic and pointer cast, new pointers can be derived from existing pointers. New pointers should inherit the metadata from the original pointers because legal pointer operations should be restricted inside the same object; thus, the derived pointer and its origin share the same intended referent. It should be noted that the address-of operator (\&) should be included in pointer arithmetic. For example, \&p[n] equals to p+n, and \&(p--\textgreater val) equals to p+offset, where the sub-field val is offset-byte from the base address of the object.

\subsection{Analysis of Saturation Access}
To analyze how SMA tolerates memory spatials errors, consider two scenarios: one with out-of-bounds reads and one without.

If there is no out-of-bounds read in the program, then all the data come from within object boundaries. When an out-of-bounds write happens, the overflowed data only affects the padding area. As the data within legal boundaries remain unchanged, the program can still get benign data for later execution. So SMA can ensure the original program behaviors and continue to execute if there is no out-of-bounds read.

From the semantic perspective, an out-of-bounds read tries to get data from nowhere. With unpredicted return values, we cannot define the outcome of later execution. So while we can prevent those illegal reads from leaking secret data, we cannot tell if later execution follows the original logic. However, if SMA does not write to one particular address more than once, the stored data is the same as the input. As a result, if out-of-bounds writes happen before an out-of-bounds read, the read can get the input as-is given a large enough padding area. In such cases, the program is likely to continue execution, but may produce meaningless outputs.

Another case is underflow. When an underflow happens, SMA changes the access address to the base address of memory object. Unlike overflows, we do not reserve paddings at the beginning of objects. This is because that most attacks rely on overflows to overwrite critical data and underflows are relatively rare. So we do not reserve paddings for underflow and leave it to the built-in error handlers in the program.

In conclusion, SMA can ensure the program to continue execution when there is no out-of-bounds read. As for the out-of-bounds reads, SMA can prevent them from leaking data, and make tha program likely to continue execution.

\section{Pointer Tagging}
Tagged pointer schemes have various advantages in managing metadata. The tagged pointer effectively correlates metadata with pointers, making the metadata query introduces almost no overhead. Also, it does not need to arrange dedicated memory space or memory operations for metadata management. However, the advantages do not come for free. Tagged pointers change the expression of pointers, which brings various challenges to both compatibility and performance. It should be noted that these challenges are shared by all the tools that use the tagged pointer scheme for metadata management, and all such tools need to find solutions for the challenges.

In this section, we will discuss the common challenges faced by tagged pointer schemes and how we solve them in SMA. Apart from the tags making addresses invalid, the common challenges can be concluded in 4 aspects. (1) Pointers may be used as integers, either by the program itself or by the compiler optimization. (2) Uninstrumented libraries are unaware of the change of pointer expression and may generate pointers without tags. (3) Attackers may corrupt the tag with arithmetic operations to manipulate the pointer metadata. (4) Microarchitecture features can affect the performance of a tagged pointer system.

\subsection{Pointers as Integers}
The C standard defines the compare operation and subtraction operation for pointers, requiring that the result is meaningful only when the two pointers point to the same object. In the real scenes, C programs usually make more assumptions to use pointers as integers, but the C standard\cite{RN262} does not guarantee those assumptions.

For the sake of the compatibility, tagged pointer systems need special care for pointer operations, i.e., removing the tag of a pointer before specific operations. If the tags remain in the pointers, certain operations generate unexpected results, possibly making the control flow deviate from the normal flow, and in the end, make the program crash or generate wrong outputs. Such errors are common in the systems where the tag of a pointer changes along with the pointer address. If all the pointers pointing to the same memory object share the same tag, then some of the pointer operations do not produce a wrong output, e.g., calculating the distance between two sub-objects with pointer subtraction.

\subsubsection{Pointer Comparisons} In the C standard, the pointer comparison produces a reasonable result only when the two pointers point to the same object, or when they are from the same aggregate object. In the real world, programs usually use pointers as integers and assume that pointers have total order, e.g., the sort of pointers. But if tags are embeded in pointers, this assumption is no longer true, especially when the tags change along with the address calculation. In SMA, a tag represents the size of an object, which does not change after the creation of the object. As a result, all the pointers to the same object share the same tag.

\subsubsection{Bitwise Operations} Generally, the operands of bitwise operations are not pointers. But some non-conforming programs use bitwise operations on pointers to detect or enforce alignment properties. In most cases, such operations should not affect the tags, because they usually manipulate the lower bits of a pointer. More carefully, we can track the pointer operand of a bitwise operation with the use-def chain in LLVM\cite{RN222}. If the result is used as a pointer later in the execution, then the tag should be preserved, or else, the tag should be removed from the result. Programs may also first use a pointer as an integer, then apply a bitwise operation on it. We can also track such cases with the use-def chain.

\subsubsection{Pointer Arithmetics} Normally, arithmetic operations apart from the addition and subtraction do not produce reasonable results on pointers. However, compilers may transform expressions involving pointers in a way that sometimes produce unexpected pointer operations. For example, compilers may transform \begin{math}(b-a)\times2\end{math} into \begin{math}b\times2+(-a)\times2\end{math} during optimization, introducing pointer multiplication operations. We can track the usage of operation’s result with LLVM's use-def chain, and decide whether or not to remove the tag.

\subsubsection{Tags Defined by Users} Some programs define their own pointer tags, either by union types or by bitwise operations. These tags contradict with the tags introduced by a tagged pointer system, and the contradiction cannot be solved without the knowledge of program semantic. A possible solution is to reserve several bits for user-defined behaviors. For example, SMA only uses the highest 6 bits of the address space, leaving 10 bits available for programmers to use.

\subsection{Coverage}
Theoretically, a defense system based on tagged pointers should tag every pointer in the program to ensure the completeness of the protection. However, it is hard to ensure complete coverage because the uninstrumented libraries cannot handle tagged pointers and may produce pointers without tags. In other occasions, we want to improve the performance by omitting the tags of pointers known to be safe. 

As a result, a robust protection system should consider how to handle the pointers without a tag. We could add extra branches to judge whether there is a tag embedded in the pointer by comparing tag bits with 0. Unfortunately, introducing extra branches affects the speed of bounds checking significantly. According to our experiments, introducing branch instructions in the bounds checking brings about 9\% performance overhead on average. As a result, we need a unified approach to handle tagged pointers and tagless pointers at the same time. In SMA, we store the complement value of the original tag in the high bits of a pointer. During the bounds checking procedure, the tags should be inverted first. For a tagless pointer, we get a tag with all 1s after the inversion, which represents a large memory object size. In this way, SMA treats tagged pointers and tagless pointers alike and avoids the low-performance branches.

The uninstrumented library is an inherent problem for every tagged-pointer-based system. Uninstrumented functions do not know the existence of the metadata tag and can generate unexpected results when using a tagged pointer. Those functions may directly use tagged pointers for memory access, compare a tagged pointer with a tagless pointer, or use a tagged pointer as an integer. Because of the unpredictable behaviors, a protection system cannot ensure the integrity of metadata tags in the uninstrumented libraries. This problem is in fact a problem of the protection boundary. A memory protection system always has a boundary. The codes within that boundary can be protected, while the codes outside of the boundary cannot. If we do not recompile the external libraries, then the protection boundary falls between the codes seen by the compiler and codes in the libraries. However, even if we instrument all of the libraries, we merely push the protection boundary from library calls to system calls.

SMA sets the protection boundary on uninstrumented library functions. When a pointer is passed into an uninstrumented function, it leaves the protection field of SMA. Defining the protection boundary requires a series of rules to guide the removal of tags and re-addition of tags. The rules for tag removal are required universally, while the re-addition rules depend on the security policy of a system.

\subsection{Metadata Corruption}\label{meta_corrupt}
Tagged-pointer-based systems should carefully distinguish pointers from integers to ensure the correct program output and protect metadata from being overwritten maliciously. Take the following scenario for an example: a program accesses the data at ptr+0x4, then attacker overwrites the offset and changes it to 0x40000000000000004. If the address calculation uses maliciously overwritten offset, the metadata tag in the result pointer is corrupted, causing the following bounds checking process to make a wrong judgment.

To avoid such metadata corruptions, we can instrument every arithmetic operation with pointer operands, and remove the metadata tag before the operation happens. Considering that pointers may be used as integers, there can be numerous such arithmetic operations, and they are hard to be distinguished from normal integer operations. Furthermore, instrumenting all of such operations brings significant performance overhead. Another method is adding hardware support. For example, the Low-Fat pointer\cite{lowfat_hard} uses dedicated instructions for pointer arithmetic. The processor checks the validity of a pointer during the execution of a pointer arithmetic instruction, and mark the pointer as invalid using hardware types when the pointer goes beyond object boundaries after the calculation.

In our prototype, we solve this problem with a pure software method. When a memory object is allocated, SMA generates a new pointer \textit{baseptr} with metadata at the same time. The new pointer \textit{baseptr} is not used in the original program execution. It always points to the base address of the memory object and is only used in the bounds-checking process. Furthermore, the \textit{baseptr} only exists in the register, unless it is swapped out by the compiler due to the lack of registers. In that case, we think that the swapping codes can be trusted because they are generated by the compiler. During the checking, SMA retrieves the base address and metadata tag from \textit{baseptr}. Because the \textit{baseptr} is transparent to the interface provided by the original program, and that attackers cannot directly manipulate the value of a specific register in our threat model, so the attacker cannot overwrite the \textit{baseptr}. In this way, we prevent the metadata from being corrupted.

The baseptr is generated according to the following rules.
\begin{itemize}
    \item For the pointer to a global variable, load from the global variable’s corresponding shadow global variable.
    \begin{IEEEeqnarray}{c}
        ptr=\&GlobalVariable\nonumber\\
        baseptr=LOAD\,(tagged\_GlobalVariable)\nonumber
    \end{IEEEeqnarray}
    \item For the pointer derived from pointer arithmetic, its baseptr is the same as the original pointer.
    \begin{IEEEeqnarray}{c}
        ptr=ptr\_ori+offset\nonumber\\
        baseptr=BASE\,(ptr\_ori)\nonumber
    \end{IEEEeqnarray}
    \item For the pointer derived from a type-cast, its baseptr is the same as the original pointer.
    \begin{IEEEeqnarray}{c}
        ptr=(type *)\,ptr\_ori\nonumber\\
        baseptr=BASE\,(ptr\_ori)\nonumber
    \end{IEEEeqnarray}
    \item For a phi node, the baseptr is one of the incoming values of the phi node.
    \begin{IEEEeqnarray}{c}
        ptr=\Phi(q_1,q_2,. . .,q_n)\nonumber\\
        baseptr=\Phi(base_1, base_2, . . ., base_n)\nonumber
    \end{IEEEeqnarray}
    \item For a pointer generated by memory allocation, calculate the tag based on the allocation size and put it at the high bits of the pointer to get baseptr.
    \begin{IEEEeqnarray}{c}
        ptr=malloc\,(size)\nonumber\\
        baseptr=tag\,(size)\ |\ ptr\nonumber
    \end{IEEEeqnarray}
\end{itemize}

When a pointer is used as a function parameter, the corresponding \textit{baseptr} should be propagated into the called function at the same time. SoftBound\cite{RN177} solve the similar metadata propagation problem by adding an extra shadow stack and change the original calling convention. For the sake of compatibility, we do not change the calling convention. Instead, we recalculate the \textit{baseptr} based on the parameter after entering the function being called. To ensure the validity of the metadata tag during the propagation, we must insert a check before the function call. If the pointer parameter is within the object boundaries before entering the function, we can still get an accurate \textit{baseptr} with the parameter.

\subsection{Hardware Support}
Most of the codes added by a tagged pointer system are simple arithmetic operations, like bitwise operation, addition, and subtraction. Although these instructions execute rather quickly in a modern processor, these extra instructions still bring pressure to the overall performance. Some architectures add MMU-based features to support the masking of tagged pointers. For example, the Address Tagging in ARMv8 architecture allows the processor to ignore the highest 8 bits during address translation\cite{RN259}, and the SPARC-M7 architecture of Oracle can support a tag as large as 32 bits\cite{RN271}. These hardware features eliminate the need for masking instructions and provide significant convenience for tagged pointer systems.

Apart from the address translation, the architecture can also add support for metadata safety, e.g., Low-Fat Pointer\cite{lowfat_hard} introduces hardware types and dedicated pointer operation instructions. Some more intrusive schemes even modify the whole processor to support the metadata management and the security policy. For example, CHERI\cite{cheri} adds dedicated registers for capability. This kind of support is generally designed for a specific protection system and binds with particular security policies. So, such support has a narrower range of applications compared with MMU-based support for masking.

\section{Implementation}
We implement the SMA prototype on x86-64 and ARMv8 architecture. The prototype is based on LLVM 10.0 and consists of two parts, LLVM C++ pass, and runtime helper functions. The project will be made open-source upon acceptance.

To make the compiler as easy to use as possible, we implement our prototype as an LLVM sanitizer. Programmers only need to add an extra flag to the original compiling procedure, and the SMA pass will be enabled and harden the program. The resulting binary can be executed as-is with no extra processing. A hardened program can automatically correct the address of a memory access upon detecting an out-of-bounds access, thus ensuring the memory accesses are always within legal boundaries and mitigating memory spatial vulnerabilities at object granularity.

\subsection{Instrumentation}
SMA instruments the program at the LLVM IR level to generate and propagate boundary metadata. The bounds checking process and the correction of out-of-bounds addresses are also inserted at IR level. The instrumentation can be divided into the following parts: adding shadow global variables, finding all the instructions to be checked, metadata generation, inserting codes for checking and correcting, and processing tags on the protection boundary.

\begin{figure}[htb]
\centerline{\includegraphics[width=0.4\textwidth]{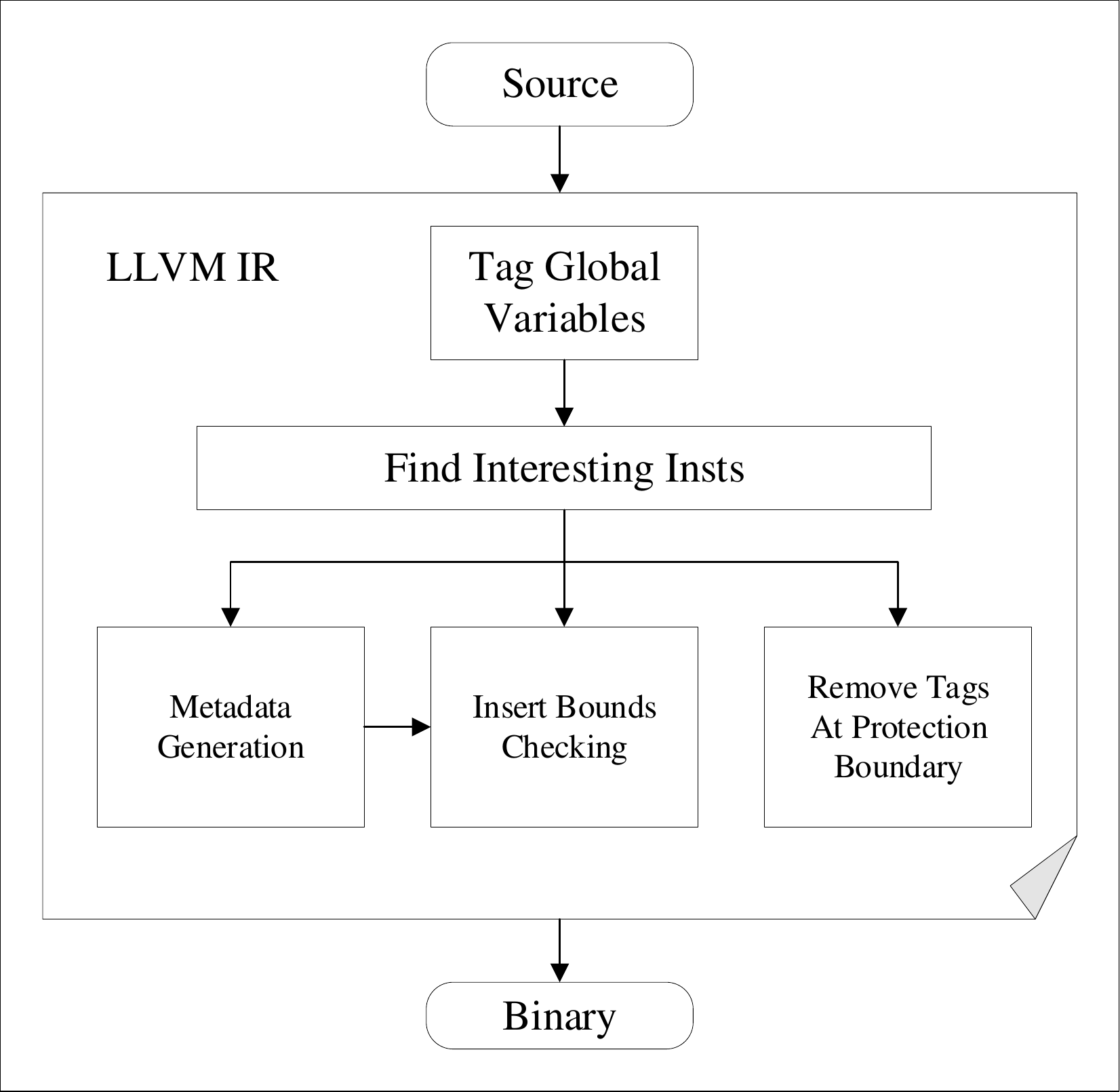}}
\caption{Instrumentation}
\label{instrumentation}
\end{figure}

\subsubsection{Global Variables}\label{gv}
As we mentioned in \ref{propagation}, the allocation of a global variable happens during the compilation. We instrument the global variables by adding a constructor at the startup of a program. This constructor generates a tagged pointer for every global variable and stores it in a shadow global variable. Shadow global varibles do not exist in the original codes. When SMA checks the validity of a global variable pointer, the corresponding shadow variable is loaded as the \textit{baseptr} for that pointer.

\subsubsection{Bounds Checking}
To check the bounds for a pointer, we need to find the positions to insert the checking codes during the compilation. Also, we need to find which objects the pointers to be checked correlate with, then generate and propagate the metadata to the checking points.

Generally, there are two ways of inserting bounds checking: checking when a pointer is changed, or checking when a pointer is dereferenced. The former monitors the pointer arithmetic and some of the integer arithmetic considering the pointer-as-integer cases, while the latter mainly monitors load and store instructions. Moreover, considering that the protection boundary of a protection system often lies on the call of functions, both of the insertion choices should pay special attention to functions with pointer parameters.

SMA adopts the latter policy, i.e., checking before the pointer dereference. In the C standard, a pointer is allowed to exceed its legal boundary on condition that the pointer is not dereferenced. So, if we check pointer validity at pointer arithmetic, then we should not correct the address at once to tolerate temporary out-of-bounds pointers. Instead, we should wait until the dereference actually happens, and this makes the checking process the same as the checking-before-dereference method. Considering the numerous cases of pointer-as-integer, we think it is more concise to monitor load and store instructions.

As stated before, we use volatile \textit{baseptr} to prevent the metadata corruption, which is generated and propagated according to the rules in \ref{meta_corrupt}. SMA first finds all the places to insert bounds checking codes, i.e., the load and store instructions. Then, SMA tracks the origin of the pointer used for memory access recursively according to the generating rules of \textit{baseptr}. This recursive process correlates the pointers being checked with the metadata of their intended referents, ensuring the correct propagation of metadata.

With the baseptr, the checking process is direct and straightforward. We can get the base and bound of a memory object with simple bitwise operations and additions after retrieving the tag with a fixed mask. Then, we only need to compare the pointer itself with the base and bound to judge whether the pointer exceeds the object boundary. In conventional out-of-bounds detection systems, an exception is thrown upon detecting the dereference of an out-of-bounds pointer. However, SMA needs to correct the address instead of simply throwing an exception. When an underflow happens, SMA changes the address to the base address of the object, and when an overflow happens, SMA changes the address to the bound address subtracting the access size. Moreover, we found that branch instructions should be avoided. According to the experiments, using branch instructions instead of cmov instructions can bring 9\% more performance overhead.

\subsubsection{Function Calls}
Similar to all the tagged-pointer-based systems, SMA has a protection boundary to decide when the tags on pointers should be removed. SMA places the boundary between the codes seen by the compiler and the codes in uninstrumented libraries, which means that SMA removes the tag when a pointer is passed to uninstrumented libraries.

Apart from uninstrumented libraries, there are some other functions that need special care. In some functions, a pointer parameter is ``passed by value'' instead of ``passed by reference''. Passing a parameter by value means that the parameter is duplicated when entering the function. As a result, we need to remove the tag of such parameters, or the tag can lead to memory access with an invalid pointer.

\subsection{Address Tagging}
The Address Tagging feature in ARMv8 architecture allows MMU to ignore the highest 8 bits of an address during translation. This feature can effectively support a tagged-pointer-based system by allowing the compiler to omit many instructions for tag removal. In SMA, we can utilize this feature and preserve tags for load/store instructions and function parameters passed by value. However, because this feature relies on the translation procedure, the use of tagged pointers beyond memory access may still bring problems; for example, the comparison between pointers still needs to remove tags. Experiments show that Address Tagging can cut more than 10\% performance overhead.

\section{Evaluation}
In this section, we evaluate the performance of SMA with MiBench\cite{mibench}, a benchmark designed for embedded environments and still widely used today. We also evaluate the security efficiency with the RIPE\cite{RN276} benchmark. Finally we use some of the real-world vulnerabilities from BugBench\cite{bugbench} to verify the ability of tolerating memory spatial errors. The performance experiments are additionally run on a 3.0GHz 16-core ARMv8 CPU with 32GB DRAM. This extra experiment utilizes the Address Tagging feature in the ARMv8 to analyze the components of performance overheads. All other experiments are run on a 1.70GHz 16-core Intel Xeon E5-2609 CPU with 32GB DRAM, and the OS is a Ubuntu 16.04 with ASLR disabled.

\subsection{Performance}
We use MiBench for performance evaluation. MiBench\cite{mibench} is a widely used embedded benchmark suite proposed in 2001. The original MiBench is relatively old, so we chose to use LLVM Testing Infrastructure\cite{llvmtest}. It is a built-in test suite of LLVM and includes MiBench. Among all of the 17 programs, only one program, typeset, does not exit with correct output. We disabled tests that run for a very short time. The disalbed tests are jpeg, ispell, stringsearch and blowfish. The runtime of these programs are all under 0.01s, making them unsuitable for performance measurements. 

We compare our performance with SoftBound\cite{RN177}. SoftBound protects both programs from both spatial errors and tmporal errors, and used by Boundless\cite{boundless} as its foundation. When compiled with SoftBound, lame, dijkstra, jpeg, typeset and rijndael cannot produce correct outputs. We omit the performance data of typeset because both SMA and SoftBound cannot run this test correctly.

\begin{table}[htbp]
	\caption{Performance Results of SMA and SoftBound on MiBench}
	\begin{center}
	\begin{tabular}{|l|l|l|l|}
		\hline
		benchmark & original & SMA & SoftBound \\ \hline
		basicmath & 641 & 649 & 590 \\ \hline
		bitcount & 116 & 117 & 225 \\ \hline
		susan & 65 & 214 & 485 \\ \hline
		lame & 227 & 735 & - \\ \hline
		dijkstra & 52 & 206 & - \\ \hline
		patricia & 162 & 228 & 233 \\ \hline
		rijndael & 75 & 101 & - \\ \hline
		sha & 28 & 76 & 120 \\ \hline
		adpcm & 434 & 480 & 622 \\ \hline
		CRC32 & 322 & 417 & 682 \\ \hline
		fft & 104 & 113 & 133 \\ \hline
		gsm & 176 & 531 & 803 \\ \hline
	\end{tabular}
	\label{performance}
\end{center}
\end{table}

\begin{figure}[htb]
	\centerline{\includegraphics[width=0.45\textwidth]{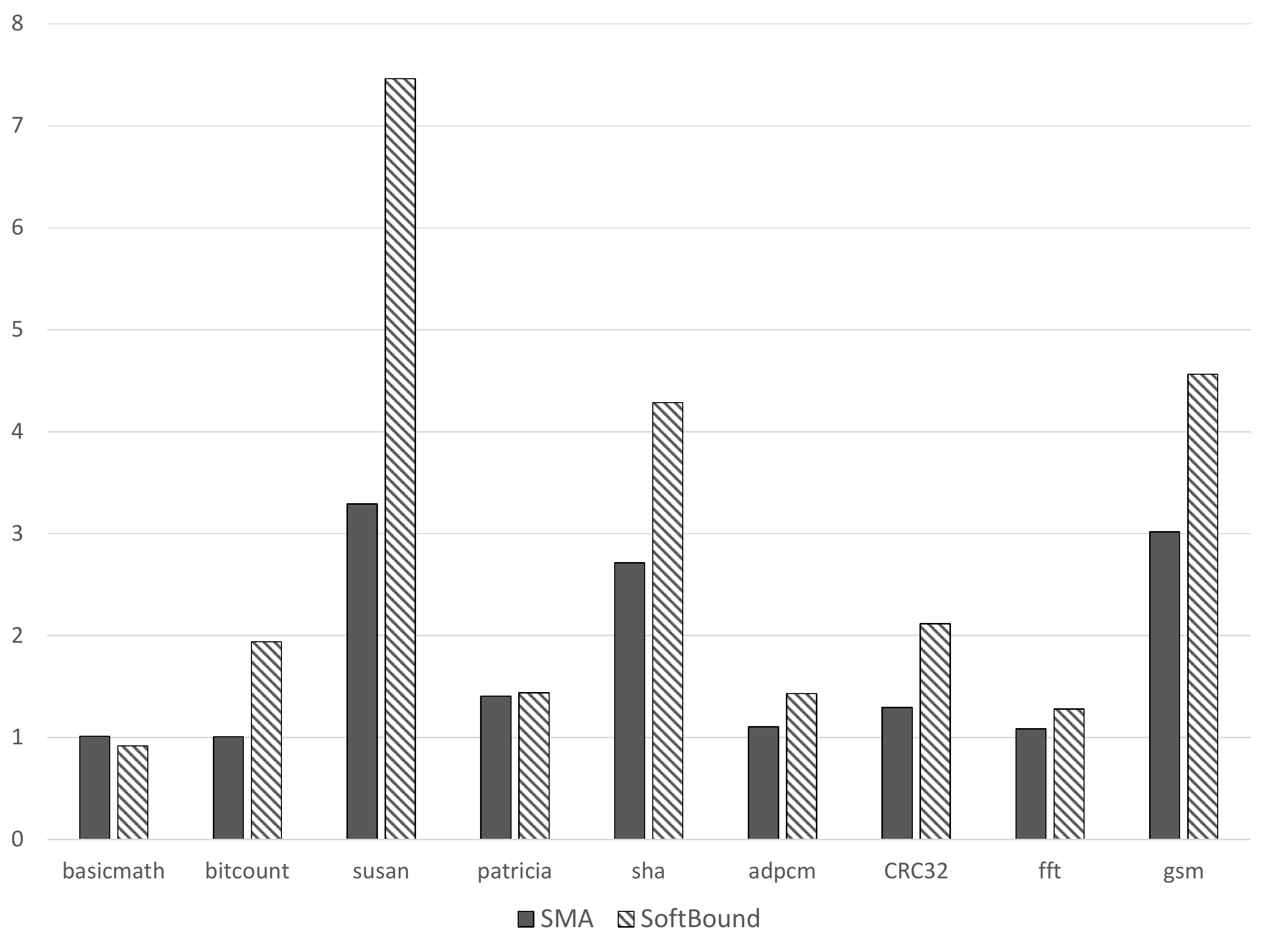}}
	\caption{Normalized Performance}
	\label{normalized-perf}
\end{figure}

Table~\ref{performance} shows the MiBench results on x86 platform. As MiBench tests all run for a relatively short time, we run every test for 1000 times and record the accumulated time. Figure~\ref{normalized-perf} shows the normalized results. From the results, we can find that SMA incurs much less overheads than SoftBound. It should be noted that SMA incurs nearly no overheads on basicmath and bitcount. This is because both tests are computation-intensive and have few memory operations. On average, SMA incurs 78\% performance overheads on MiBench.

\subsection{Overheads Analysis}
We further analysis the components of overheads on ARM platform. The Address Tagging feature of ARMv8 allows us to omitting the tag removal instructions for loads and sotres. We analyze overheads with SPEC CPU2017, which contains various complex real-world programs, including video encoding and decoding, simulation, and artificial intelligence. Tests in SPEC CPU2017 are much larger in scale than those in MiBench. The larger scale brings more memory accesses and amplifies the impact of each component of the overall overhead so that it can be observed more clearly.

We chose the 12 pure C/C++ programs in the speed set of SPEC CPU2017 for the analysis. Among the 12 programs, 600.perlbench\_s, 602.gcc\_s, 623.xalancbmk\_s, 641.leela\_s cannot produce correct outputs. Several similar bounds-checking tools reported problems alike. 600.perlbench\_s and 602.gcc\_s cannot execute correctly without modifications to the source code. As reported by SGXBounds, gcc has unions of pointer-ints and manipulates high bits of the pointer. For 623.xalancbmk\_s and 641.leela\_s, they encounter issues in the uninstrumented library codes by comparing tagged pointers with tagless pointers.

\begin{table*}[htbp]
	\caption{SPEC CPU2017 Results on ARM}
	\begin{center}
	\begin{tabular}{|l|l|l|l|l|l|l|l|}
		\hline
		\multicolumn{1}{|c|}{benchmark} & \multicolumn{1}{c|}{llvm-base} & \multicolumn{1}{c|}{\begin{tabular}[c]{@{}c@{}}with\_mask\\ with\_check\end{tabular}} & \multicolumn{1}{c|}{\begin{tabular}[c]{@{}c@{}}no\_mask\\ with\_check\end{tabular}} & \multicolumn{1}{c|}{\begin{tabular}[c]{@{}c@{}}no\_mask\\ no\_check\end{tabular}} & \multicolumn{1}{c|}{\begin{tabular}[c]{@{}c@{}}with\_mask\\ no\_check\end{tabular}} & \multicolumn{1}{c|}{\begin{tabular}[c]{@{}c@{}}no\_mask\\ with\_check\\ branch\end{tabular}} & \multicolumn{1}{c|}{\begin{tabular}[c]{@{}c@{}}with\_mask\\ with\_check\\ branch\end{tabular}} \\ \hline
		mcf\_s & 1701 & 3439 & 3365 & 1732 & 2101 & 3766 & 3692 \\ \hline
		omnetpp\_s & 1011 & 2364 & 2263 & 1576 & 1753 & 2384 & 2378 \\ \hline
		x264\_s & 744 & 2322 & 2118 & 817 & 1136 & 2513 & 2516 \\ \hline
		deepsjeng\_s & 634 & 1570 & 1396 & 791 & 904 & 1664 & 1651 \\ \hline
		xz\_s & 4304 & 6977 & 6534 & 4442 & 4844 & 7268 & 7062 \\ \hline
		lbm\_s & 3217 & 3260 & 3641 & 3840 & 3023 & 3763 & 3597 \\ \hline
		imagick\_s & 14492 & 30541 & 27934 & 14772 & 16922 & 31765 & 31719 \\ \hline
		nab\_s & 6046 & 9148 & 8663 & 6077 & 6564 & 8994 & 9014 \\ \hline
	\end{tabular}
	\label{spec}
	\end{center}
\end{table*}

Table~\ref{spec} shows the SPEC CPU2017 results on the ARM platform. In the table, ``no\_mask'' means omitting the removal of pointer tags with Address Tagging, and ``no\_check'' means omitting the bounds checking procedures. The last two columns use branch instructions instead of cmov instructions to implement the bounds checking procedure. From the results, we can find that the removal of pointer tags makes up 11-16\% of the total overhead, so the Address Tagging feature is effective in improving the overall performance.

\begin{figure}[htb]
	\centerline{\includegraphics[width=0.5\textwidth]{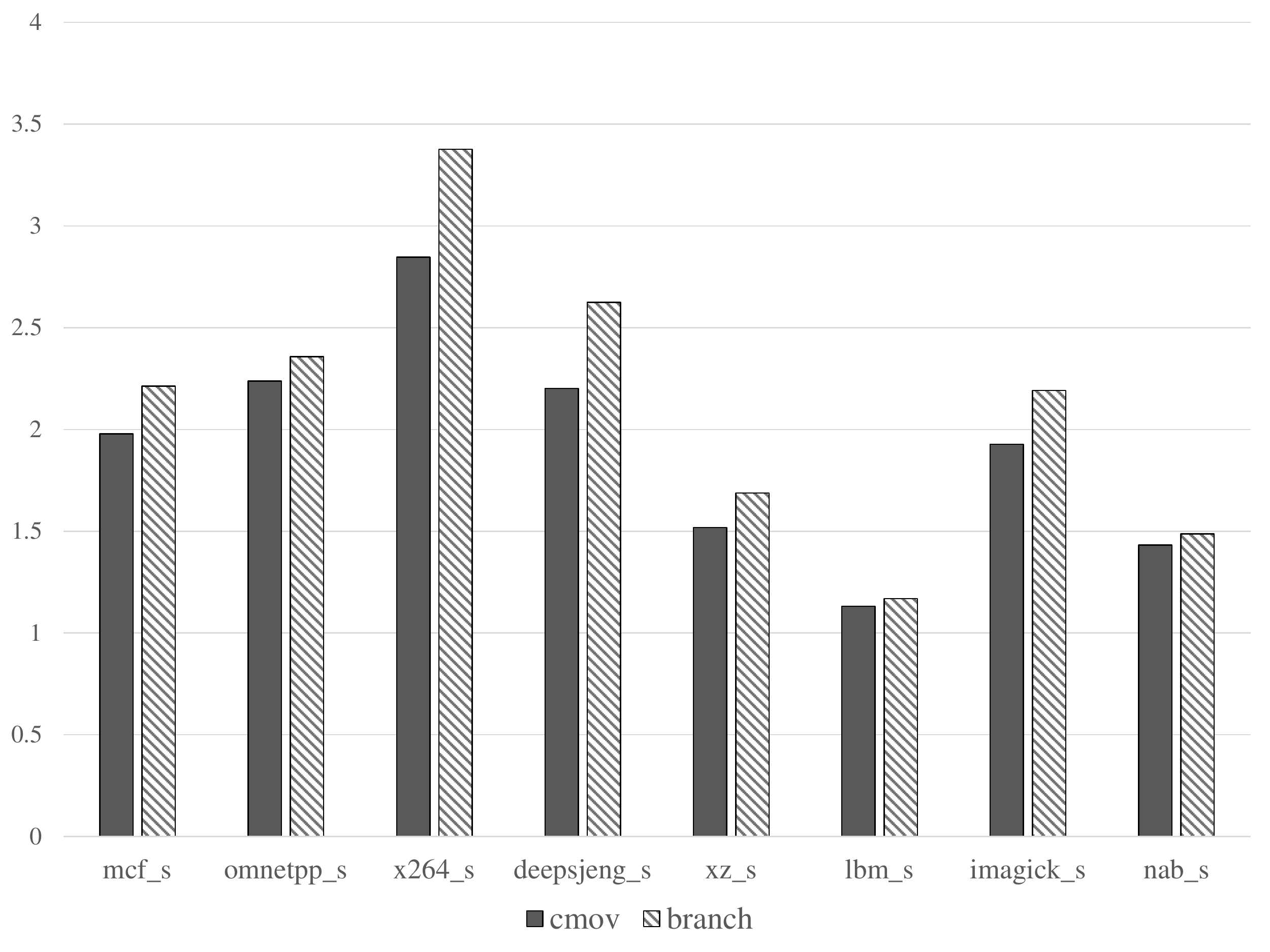}}
	\caption{Normalized Execution Time of cmov/branch}
	\label{cmov_branch}
\end{figure}

\begin{figure}[htb]
	\centerline{\includegraphics[width=0.5\textwidth]{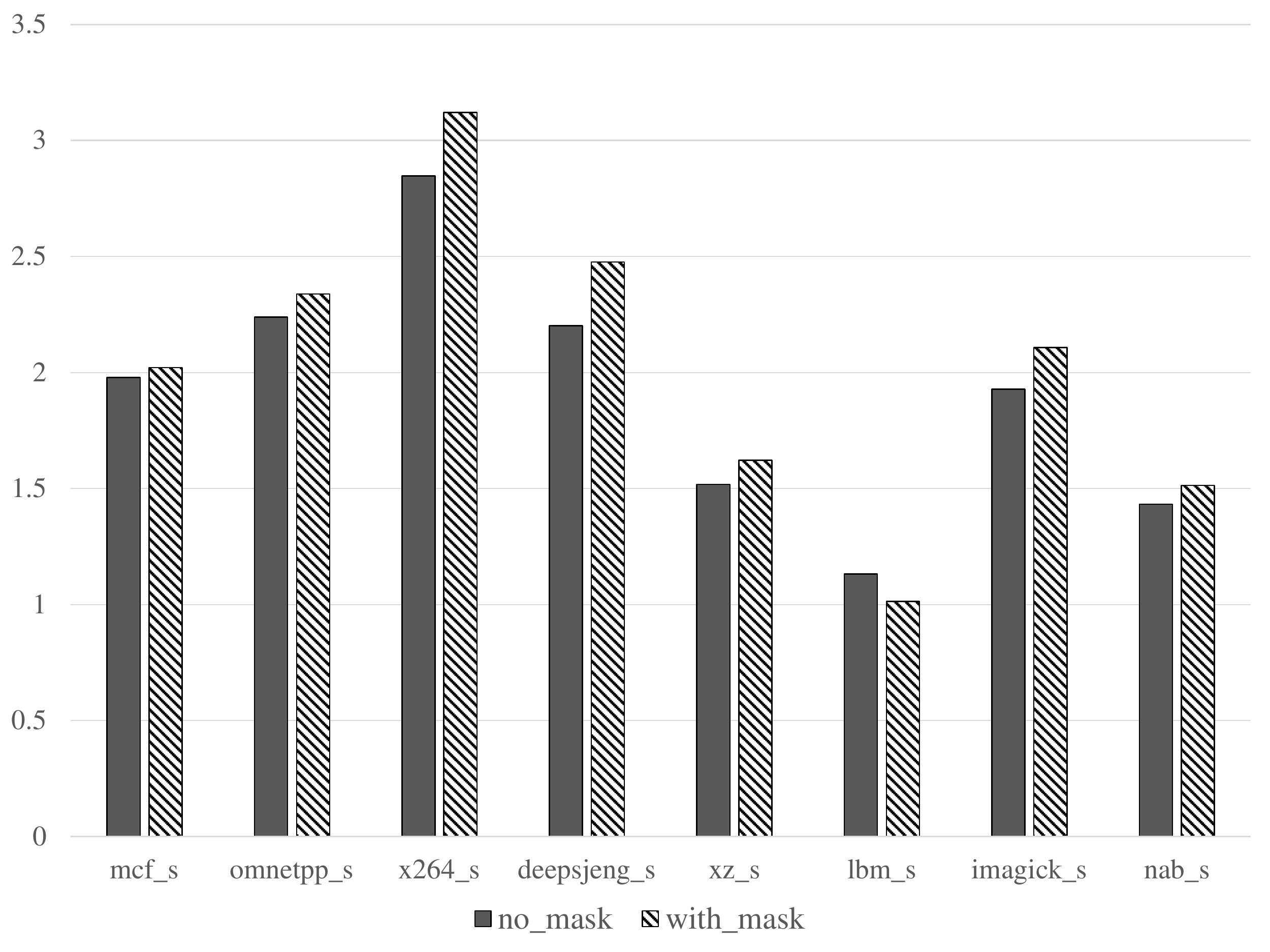}}
	\caption{Normalized Execution Time with/without Masking}
	\label{masking}
\end{figure}

As stated before, we found that the branch instructions would slow the checking process, and we did further research on this subject. As shown in Fig~\ref{cmov_branch}, the branch instructions incur about 9\% more execution time. Moreover, the benchmarks in which branch instructions incur the most overheads are also the ones the Address Tagging can benefit most. In Fig~\ref{masking}, we can observe that x264\_s, deepsjeng\_s, and imagick\_s benefit most from the Address Tagging feature, and they also suffer most from the branch instructions. This correlation originates from the fact that SMA inserts the checking process before the dereference of pointers, so the number of checking process is the same as the number of the tag removal process.

\begin{figure}[htb]
	\centerline{\includegraphics[width=0.35\textwidth]{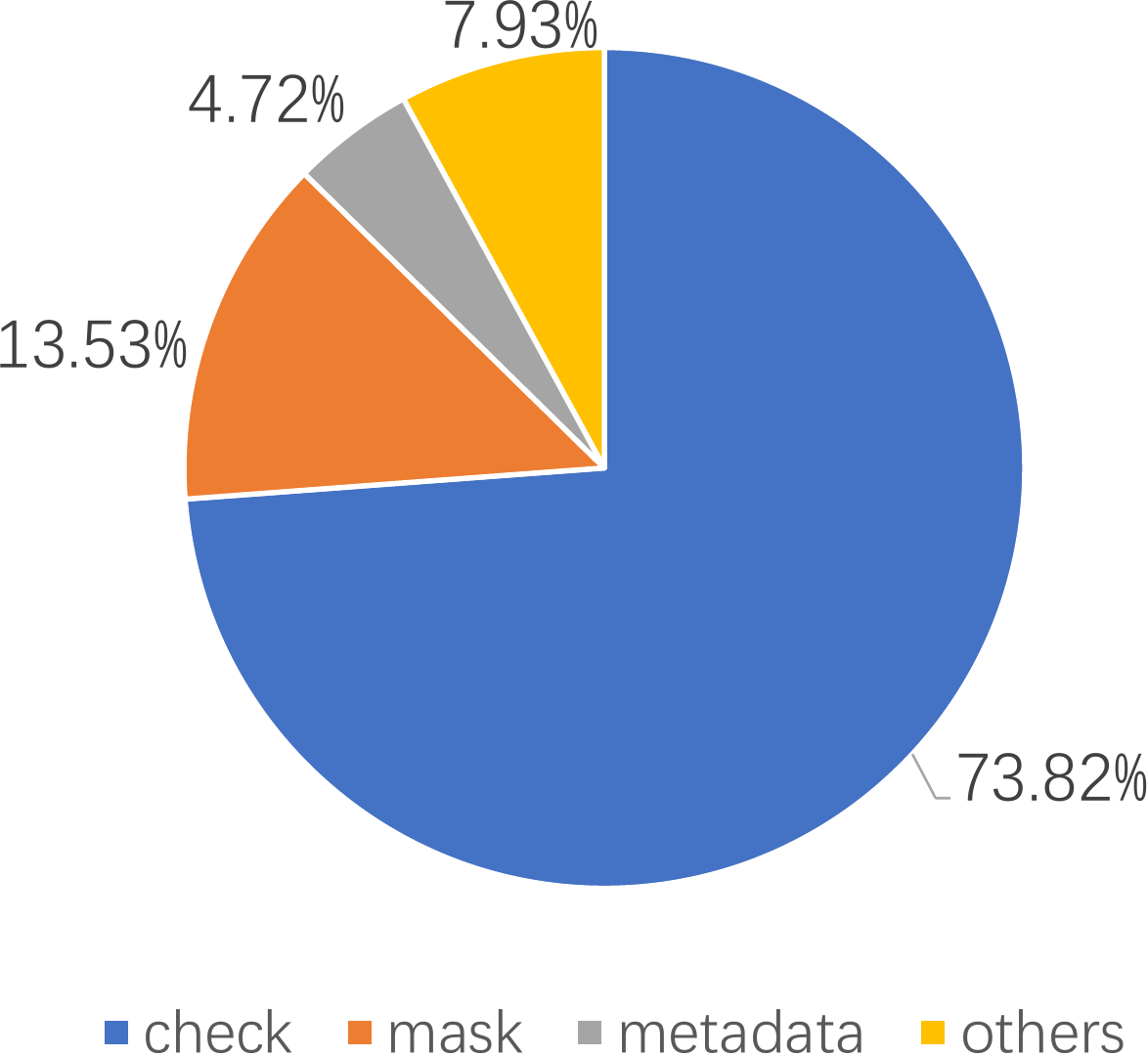}}
	\caption{Components of the Overheads}
	\label{overheads}
\end{figure}

The overheads of SMA consist of three main parts, checking, masking, and metadata management. The total overhead is defined as the overhead when the masking and checking are both enabled. As shown in the Fig~\ref{overheads}, the checking process makes up most of the overheads in both encoding schemes. In the floating encoding, the checking process incurs more overheads than that in the buddy encoding because the floating encoding needs more instructions to decode tags. Moreover, adding instructions to a program brings extra pressure to the execution pipeline, and may affect the branch prediction or other microarchitecture level behaviors. We classify such overheads to the ``others'' category in Fig~\ref{overheads}. The details of this part of overheads require in-depth experiments, and we leave them for future work.

\subsection{Security Efficiency}
We use the RIPE\cite{RN276} benchmark for security efficiency evaluation. RIPE is an extension of Wilander’s and Kamkar’s testbed and aims to provide a standard way of testing the coverage of a defense mechanism again memory spatial errors. The RIPE testbed has 5 dimensions: location, target code pointer, overflow technique, attack code, and function abused. It covers 850 attack forms, and if attacks with different payloads are viewed as different attack forms, RIPE can generate over 2000 attack forms.

We compare the security efficiency of original Clang, SoftBound-CETS\cite{RN177} and SMA on the x86-64 platform with system ASLR disabled. There are 1334 attack forms considered possible by RIPE, and each of the possible attacks was tried for 3 times. The results of the attack experiment are listed in Table~\ref{tab1}. In the table, ``partly'' means some of the tries succeeded while the others failed.

\begin{table}[htbp]
	\caption{RIPE results}
	\begin{center}
			\begin{tabular}{|l|l|l|l|l|}
				\hline
				& Success & Partly & Fail & Fail\%  \\ \hline
				\textbf{Clang}     & 1289    & 13     & 32   & 2.40\%  \\ \hline
				\textbf{SoftBound-CETS} & 0     & 0      & 1334 & 100\% \\ \hline
				\textbf{SMA}        & 65      & 0      & 1269 & 95.13\% \\ \hline
		\end{tabular}
		\label{tab1}
	\end{center}
\end{table}

From the results, we can find that SoftBound-CETS blocked every possible attack. SoftBound-CETS is a complete defense mechanism mitigating both spatial errors and temporal errors\cite{RN9}. As for SMA, there are some attacks that succeeded. Further investigation of the results showed that the successful attacks breach the protection through sub-object overflows. For now, our prototype does not support saturation accesses on sub-objects, but it still mitigated most of the memory spatial errors and greatly increased the security efficiency comparing to the original Clang. It should be noted that none of the failed attacks reported segmentation fault, so all the failure of attacks should be attributed to the saturation access mechanism instead of compatibility issues. From the results, we can conclude that SMA is an effective protection against the memory spatial error.

\subsection{Tolerance}
There are 6 programs with buffer overflow vulnerabilities in BugBench, namely bc, gzip, man, ncompress, polymorph and squid. Some of the bugs cannot be reproduced with modern compilers and libraries. For example, poly morph 0.4.0 in BugBench reports an buffer overflow error during compilation, and the overflow in gzip 1.2.4 is caught by a system call. We reproduced the vulnerabilities in bc 1.06 and ncompress 4.2. Both of the programs will trigger a segmentation fault when provided with a buggy input. We recompiled the two programs, and they both exited normally with error logs generated by built-in error handler.  

\section{Discussion}
\subsection{Bounds Narrowing} Our prototype does not detect sub-object overflow for now out of compatibility consideration. Because SMA correlates metadata with pointers instead of objects, it is possilbe for SMA to do bounds narrowing and store sub-object metadata in a pointer. To support bounds narrowing, we need to add type layout information into the program. EffectiveSan\cite{effectivesan} provides a way to achieve this. However, maintaining metadata for every sub-object often introduces intrusive modifications to program source codes for SMA, e.g., the memory layout of a structure may have to be changed to fulfill the alignment requirements, which may seriously harm the compatibility of the program. We leave the integration of type information as a future work.

\subsection{Protection Boundary} Currently, we place the protection boundary between the codes seen by the compiler and the codes of uninstrumented libraries. As stated before, we can push this boundary to the system calls if we recompile the library codes. To further expand the protection boundary, we need to modify system ABIs and kernel codes. CHERI has accomplished part of that modification and has made system ABIs aware of the CHERI capability\cite{cheriabi}. In SMA, the size of the pointers does not change, so such modification would be less intrusive when expanding the protection boundary to OS kernel, or even BIOS.

\subsection{Optimizations} Fig~\ref{overheads} shows that the overheads of SMA mainly come from the checking process. There are serval methods to reduce the number of chekings, such as ABCD\cite{abcd} and WPBOUND\cite{wpbound}, which can be integrated with SMA. Also, if we could modify the hardware, then we are able to add more dedicated support for SMA. For example, we could integrate the decoding of tags with the execution of load/store instructions. Those instructions inherently have a long execution time because of the interaction with another memory hierarchy. So, adding a checking process consisted of only arithmetic operations should not expand the critical path, and the overhead of checking can be easily overlapped. We believe that adding support at the ISA level for bounds-checking is a promising topic for future research.

\section{Related Works}
\subsection{Tolerance} The idea of tolerating the memory errors instead of terminating the program is much like Boundless\cite{boundless} and failure-oblivious computing\cite{failureob}. The main difference between SMA and these two mechanisms is the way of handling memory errors. Instead of discarding or storing all the out-of-bounds data, SMA removes the complicated linked list in Boundless while still reserves parts of the input data. Moreover, Boundless is built on SoftBound and failure-oblivious is built on J\&K\cite{RN204}, both of them are known to increase great performance overheads.

\subsection{Tagged Pointers} DeltaPointer\cite{RN250} stores the distance between current address and the upper bound of an memory object. It moves the bounds checking procedure to MMU and greatly reduces the performance overheads. But DeltaPointer can only detect overflows and leaves underflows undetected, which is its main drawback.

SGXBounds\cite{RN249} also adopts tagged pointers. SGXBounds aims to protect programs within the SGX environment. SGX uses 64-bit pointers but only allows porgrams to use a 32-bit address space. SGXBounds stores a pointer to the upper bound of a memory object in the top 32 bits of a 64-bit pointer, and stores the base address next to the end of that object. The application of SGXBounds is restricted in SGX, and retriving the base address will introduce an extra memory load.

\section{Conclusion}
This paper proposes a novel, automatic memory spatial error mitigation system with low memory and performance overhead. We use tagged pointers to record the boundary metadata of memory objects and require special alignment during memory allocation to quickly locate the pointer’s intended referent. With Saturation Memory Access, we can restrict all the memory accesses within the boundaries of objects. The prototype is implemented on LLVM 10.0 and is able to mitigate nearly all of the attack forms in the RIPE benchmark, incurring 78\% performance overhead on MiBench. Furthermore, with hardware supports such as Address Tagging in ARMv8, we can achieve even lower performance overhead. To be concluded, SMA can stop most of the attacks caused by memory spatial errors with competitive overhead. To foster further research in the field, we will make the source code of our SMA prototype available as open-source upon acceptance.

\bibliographystyle{IEEEtran}
\bibliography{ref}

\end{document}